\documentstyle[epsfig,prb,aps]{revtex}

\newcommand{\be}{\begin{equation}}
\newcommand{\ee}{\end{equation}}

\begin{document}
\title{\bf Size-effects in the Density of States in NS and SNS junctions \\
}

\author{M. Blaauboer, R.T.W. Koperdraad, A. Lodder and D. Lenstra} 

\address{Faculteit Natuurkunde en Sterrenkunde, Vrije Universiteit,
         De Boelelaan 1081, 1081 HV Amsterdam, The Netherlands}

\maketitle

\begin{abstract}
\normalsize{The quasiparticle local density of states (LDOS) is studied
in clean NS and SNS junctions with increasing transverse size, from quasi-one-dimensional to three-dimensional. It is shown that finite transverse dimensions are related to pronounced effects in the LDOS, such as fast oscillations superimposed on the quasiparticle interference oscillations (for NS) and additional peaks in the bound state spectrum in the subgap region (for SNS). Also, the validity of the Andreev approximation is discussed. It turns out
to be an acceptable approximation in all situations tested.}
\end{abstract}

\pacs{PACS numbers: 74.80.Fp 
        {\tt cond-mat/9605078}
}


\section{Introduction}

In 1964 Andreev described a new kind of reflection process by which electrons incident on a Normal-Metal Superconductor (NS) interface are reflected
as holes, and vice versa\cite{andreev}. This process, now known as Andreev reflection, led shortly after its discovery to both theoretical and experimental work on tunneling transport and the related local density of states (LDOS) in small superconducting structures involving at least one NS interface. 
The dimensions of the samples perpendicular to the current flow were
essentially macroscopic and the corresponding theories \cite{mcmillan,ishiidos}
were three-dimensional (3D).

Miniaturization of devices led to the development of mesoscopic
physics \cite{beenakker1}. The initial model approaches in that new
field were one-dimensional (1D). Attention was focused on an obviously
nonequilibrium property, the conductance, and guided by
Landauer's early result for it \cite{landauer,buttiker}.
Finite transverse dimensions were considered by counting the
number of transverse modes. The extension of Landauer's formula
to the NS system was given by Lambert \cite{lambert}. In these
studies only the total density of states of 1D systems enters, being
inversely proportional to the velocity. In calculating the
Josephson current in SNS junctions Beenakker \cite{beenakker2}
applied a more advanced expression for the total density of states
given by Akkermans et al \cite{akkermans}.

As far as the LDOS is concerned, even recent studies \cite{furusaki,tanaka,kiesplehn} are 3D as yet. In this paper we calculate the LDOS of NS and SNS junctions with finite transverse dimensions, by this considering effectively 1D systems and all 
possibilities between 1D and 3D. A Green    
function approach\cite{proefschrift,rutger}
is used, inspired by Ishii\cite{ishii} and Tanaka and Tsukada\cite{tanaka}.

In case of the NS junction, we first investigate the LDOS in the quasi-1D limit of this junction. "Quasi-1D" means in the limit of transverse system size going to zero. The LDOS is shown to exhibit oscillations as
a function of both energy $E$ and distance from the interface. This result reproduces previously observed  and analyzed oscillations in tunneling experiments\cite{rowell,tomasch}. 
We then increase the transverse dimensions and find the appearance of
additional oscillations. In progressively refined
applications of Scanning Tunneling Microscopy \cite{crommie} 
these oscillatory effects in the LDOS might well become detectable 
in the near future. In the case of infinite transverse dimensions the 
additional oscillations disappear again.

In the SNS junction, we study the LDOS in the normal region for energies below the superconducting gap, and find in the quasi-1D junction one bound state. With increasing transverse system size, the number of bound states increases. 

A point of discussion in our analysis is the role of the Andreev approximation, which is demonstrated to be a good approximation for both the NS and the SNS junctions. Its effect becomes noticeable for large transverse dimensions only.

In section \ref{sec:theory} we give a brief outline of the theory and the model used. The LDOS in NS and SNS junctions is discussed in sections \ref{sec:NS} and
\ref{sec:SNS}, followed by some conclusions in section \ref{sec:concl}.

\section{Theory}
\label{sec:theory}

The Green function method used in this paper is described in Ref.13, and
will be published in a forthcoming publication \cite{rutger}. We refer the reader to these papers for an extensive description of it,
and here only summarize the aspects which are of direct importance for the calculation of the local density of states. 

The Green function describes the various ways of propagation from one point in space ${\bf r}$ to another one ${\bf r}^{'}$. Here we study clean metallic systems consisting of a few layers, in which scattering only takes place at the
interfaces between the layers. In the presence of an interface, the total Green function $G({\bf r}, {\bf r}^{'})$ usually consists of two terms: one bulk term, accounting for propagation in the material without any influence from the interface, and a scattering contribution from interaction with the interface. We use the expressions for the homogeneous bulk superconductor as given by Ishii\cite{ishii} and follow Koperdraad {\it et al.}\cite{proefschrift,rutger}
in determining the scattering matrix elements for two simple systems: a planar
NS junction with only one interface, and a SNS junction containing two interfaces.

The central quantity of this paper, the local quasiparticle density of states
in 3D inhomogeneous superconducting structures, is calculated from the
matrix Green function corresponding to the Bogoliubov equations for quasiparticle states\cite{degennes}. This Green function is the solution of the following matrix equation
\be
\left[ i\omega_{n} \tau_{0} - K\tau_{3} - D(x) \right] G({\bf r}, {\bf r}^{'})
= \delta({\bf r} - {\bf r}^{'})\tau_{0}
\label{eq:Bog3D}
\ee
where $\omega_{n}= (2n+1)\pi k_{B} T$ are the Matsubara frequencies, $K$ is the free particle Hamiltonian minus the chemical potential $\mu$, in atomic units ($\hbar$=$2m$=1) given by
\[
K = - {\bf \nabla}^2 - \mu,
\]
and the matrices $\tau_{0}$, $\tau_{3}$, and $D(x)$ are given by
\begin{displaymath}
\tau_{0} = \left[ \begin{array}{cc}
1 \  & \   0 \\ 
0  \ & \  1 
\end{array} \right],\ 
\tau_{3} = \left[ \begin{array}{cr}
1 \  &   0 \\ 
0 \  &   -1 
\end{array} \right],\ 
D(x) = \left[ \begin{array}{cc}
0   &   \Delta(x) \\ 
\Delta^{*}(x)   &   0 
\end{array} \right].
\end{displaymath}
Here $\Delta (x)$ denotes the superconducting pair potential, which is zero in the normal part of the system.
\vspace{0.5cm}

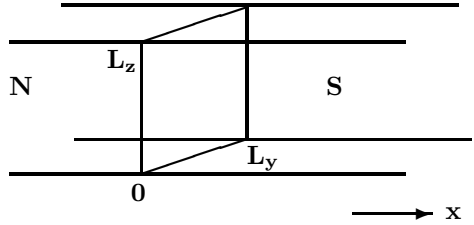
\begin{figure}[t]
$$\begin{picture}(200,80)
\thicklines
\put(0,0){\line(1,0){150}}
\put(0,50){\line(1,0){150}}

\put(50,0){\line(0,1){50}}

\put(50,0){\line(3,1){40}}
\put(50,50){\line(3,1){40}}

\put(90,13){\line(0,1){50}}

\put(20,64){\line(1,0){150}}
\put(25,13){\line(1,0){150}}

\put(46,-10){\bf 0}
\put(0,30){\bf N}
\put(120,30){\bf S}
\put(37,40){\mbox{\boldmath $\bf L_{z}$}}
\put(90,4){\mbox{\boldmath $\bf L_{y}$}}

\put(130,-15){\vector(1,0){30}}
\put(165,-17){\bf x}

\end{picture}$$
\vspace*{5mm}
\caption{ \protect{\label{fig:NSjunction}}
3D planar NS junction with finite transverse dimension $L_{y}L_{z}$}
\end{figure}

We apply the above sketched formalism to systems which are infinitely long in the x-direction, but of finite length $L_{y}$ and $L_{z}$ in the transverse y- and z-directions, for example to a NS junction with a square transverse cross section $L_{y}L_{z}$ as drawn in FIG.~\ref{fig:NSjunction}.
Contrary to the translationally invariant situation treated usually\cite{tanaka,proefschrift}, in which $G({\bf r}, {\bf r}^{'})$ depends on the differences $y - y^{'}$ and $z - z^{'}$ only, in the present case the dependence on $y$, $z$
and the primed coordinates is not reduced to differences. This means that
variations in the LDOS in the transverse directions survive after taking $y^{'}=y$ and $z^{'}=z$, the latter substitutions being required in calculating
the LDOS. However, such dependence cannot be measured and further it depends on the precise preparation of the boundaries. Therefore it is sufficient to take the average over the transverse directions. This can be done as follows. First, the boundary conditions are chosen such that the Green function vanishes at the boundaries in both transverse directions. Only the functions contribute that are proportional to sin($k_{y}y$) and sin($k_{z}z$), with $k_{y} = \frac{n_{y} \pi}{L_{y}}$ and $k_{z} = \frac{n_{z} \pi}{L_{z}}$, $n_{y}$ and $n_{z}$ being non-negative integers. The expansion of the full Green function in terms of these functions has expansion coefficients $G(x,x^{'},k_{y},k_{y}^{'},k_{z},k_{z}^{'})$. Subsequently, one puts $y^{'}=y$
and $z^{'}=z$ in this expansion and averages it over the wire's cross section. In this way, only terms with $k_{y}^{'}=k_{y}$ and $k_{z}^{'}=k_{z}$ survive, the corresponding expansion coefficients of which are denoted by $G(x,x^{'},k_{y},k_{z},i\omega_{n})$. The variable $i\omega_{n}$ is added because the Green function is still a solution of Eq. (\ref{eq:Bog3D}). Finally one manipulates Eq. (\ref{eq:Bog3D}) according to the expansion and averaging procedure indicated above and one finds that it reduces to
\be
\left[ i\omega_{n} \tau_{0} - K_{x}\tau_{3} - D(x) \right]\,  G(x,x^{'},k_{y},k_{z},i\omega_{n}) = \delta(x - x^{'})\tau_{0}
\label{eq:Bog1D}
\ee
with
\be
K_{x} = - \frac{d^2}{dx^2} - k_{F_{x}}^2 \ \ \mbox{\rm and} \ \ k_{F_{x}}^2 = 
\mu - k_{y}^2 - k_{z}^2
\ee
Using properly normalized functions in the complete sets in the $y$ and $z$ directions, it is found that the quasiparticle LDOS\cite{proefschrift} reduces to
\be
\rho(x,E) = - \frac{1}{\pi}\, \lim_{\delta \rightarrow 0}\, \frac{1}{L_{y} L_{z}} \sum_{k_{y}, k_{z}} \mbox{\rm Im}\, G_{11} (x,x,k_{y}, k_{z}, E +  i\delta)
\label{eq:ldos}
\ee

$G_{11}$ is the upper left matrix element of $G(x,x^{'},k_{y},k_{z},i\omega_{n})$ with $x = x^{'}$,
and the standard replacement of $i\omega_{n}$ by $E + i\delta$ has been applied\cite{tanaka}. Im $G_{11}$ denotes the imaginary part of $G_{11}$ and $E$ is the quasiparticle energy measured with respect to the Fermi energy $\mu$. 
Eq. (\ref{eq:ldos}) forms the basis of all our present calculations. 

A well-known and often applied approximation in calculations concerning inhomogeneous superconducting structures is the so-called Andreev approximation (AA). It was first introduced by Andreev\cite{andreev}
and can be stated in several ways. Perhaps the simplest is to say that in AA normal reflections due to mismatch of wavevectors at the normal and superconducting sides of a NS interface are neglected. Mathematically, it means
that we make a series expansion of the electron and hole wavevectors, and only take into account terms up to first order in $E/\mu$ and $\Delta/\mu$.
If the wavevector appears as a prefactor, it is approximated even further and taken to be the Fermi wavevector.\\
In 1D systems, where one is often interested in energies $E$ deviating very little from the large Fermi energy, this is regarded as a good approximation.
In our 3D systems, with an effective chemical potential $k_{F_{x}}^2 = \mu - k_{y}^2 - k_{z}^2$, application of the AA relies on the assumption that $E, \Delta \ll k_{F_{x}}^2$.
It is however not a priori clear whether this assumption is valid for all $k_{y}$ and $k_{z}$; especially for large transverse dimensions $L_{y}$ and $L_{z}$, when $\rho$, as given by Eq. (\ref{eq:ldos}), is a sum over many wavevectors $k_{y}$, $k_{z}$, there are terms for which $k_{F_{x}}^2$ is of the same order of magnitude as $E$
and $\Delta$.

\section{The NS junction}

\label{sec:NS}
We consider a normal-metal to superconductor junction as in FIG.~\ref{fig:NSjunction}. From now onwards, in the actual calculations the transverse dimensions are taken equal, so $L_{y} = L_{z} \equiv L_{t}$. In principle, the pair potential $\Delta$ has to be determined self-consistently\cite{kiesplehn}, but as a first approximation
we take it to be zero in the normal region and constant in the superconductor. So the proximity effect is not included.
The chemical potentials of the normal metal and superconductor are
denoted by $\mu_{N}$ and $\mu_{S}$ respectively.

\begin{figure}
\vspace*{5mm}
\centerline{ \epsfig{figure=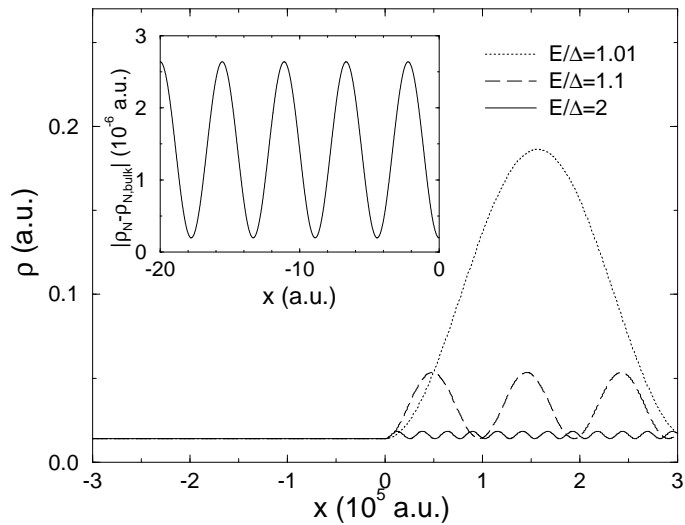,width=9cm}}
\vspace*{5mm}
\caption{ \protect{\label{fig:quasi1D}}
LDOS as a function of distance $x$ in a quasi-1D NS junction with $L_{t}$=4; $\mu_{N}$=$\mu_{S}$$\equiv$$\mu $=0.5 Ry and $\Delta$=0.0001 Ry. The interface is located at x=0. The inset shows Friedel oscillations in $| \rho_{N} - \rho_{N,bulk}| \cdot 10^{-6}$ as a function of $x$ in the normal metal for $E/\Delta$=1.01; note that the scales on the y-axis differ by a factor $10^{5}$.
}
\end{figure}

FIG.~\ref{fig:quasi1D} shows the quasi-1D LDOS in a NS junction with $\mu_{N} = \mu_{S} \equiv \mu$, as a function of $x$ for various energies $E> \Delta$. The calculation is exact, i.e. without the AA, and typical values of the chemical potentials and the gap energy in a superconductor are used, expressed in atomic units.
Oscillations are clearly
visible, both in the normal metal LDOS ($\rho_{N}$, see inset) and in the superconductor LDOS ($\rho_{S}$). The oscillations in $\rho_{N}$ are the well-known Friedel oscillations, due to interference of incident and 
reflected electron wave functions, which give rise to a component in
$\rho_{N}$ proportional to ${\rm cos}(2 k^{e} x)$, where 
$k^{e} = \sqrt{\mu_{N} + E}$.
The characteristic wavelength of the oscillations is then $L_{N}^{\rm char}
\equiv \frac{2 \pi}{2 k^{e}} \approx 4 $. Since there is no potential barrier at the interface, the amplitude of these oscillations is very small,
5 orders of magnitude smaller than the ones in the superconductor.
These Friedel oscillations would not be found in the AA. 

The oscillations in $\rho_{S}$ are caused by quasiparticle
interference. Mc Millan and Rowell\cite{rowell} named them the superconducting analogue of the Friedel oscillations. Let the electronlike (holelike) quasiparticle wavenumber be denoted by $q^{e}$ ($q^{h}$); the oscillatory component in $\rho_{S} \sim {\rm cos} ((q^{e} - q^{h})x) \approx {\rm cos} (\sqrt{E^2 - \Delta^2}x / \sqrt{\mu_{S}})$ then gives rise to a characteristic wavelength $L_{S}^{\rm char} \equiv \frac{2 \pi \sqrt{\mu_{S}}}{\sqrt{E^2 - \Delta^2}} \gg L_{N}^{\rm char}$.

\begin{figure}
\vspace*{5mm}
\centerline{\epsfig{figure=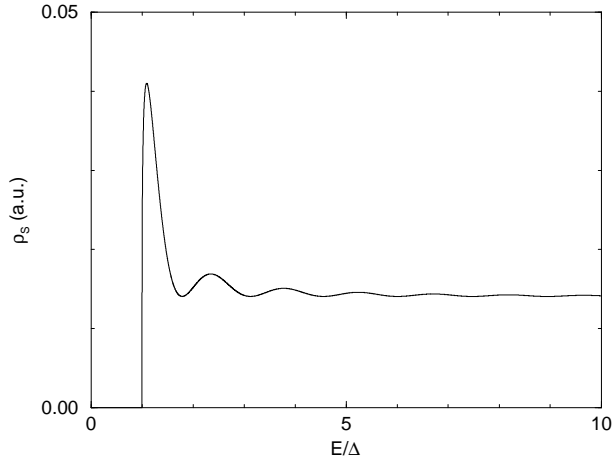,width=9cm}}
\vspace*{5mm}
\caption{ \protect{\label{fig:Friedelbekend}}
LDOS as a function of $E/\Delta$ in a quasi-1D NS junction at $x=3\cdot10^{4}$ in the superconductor.  $\mu_{N}$=$\mu_{S}$$\equiv$$\mu $=0.5 Ry,  $\Delta$=0.0001 Ry and $L_{t}=4$.
}
\end{figure}
For completeness, we also plot the oscillations in $\rho_{S}$
as a function of $E/\Delta$ for a fixed position in the superconductor, see FIG. ~\ref{fig:Friedelbekend}. They were measured in thin films by Rowell and Mc Millan\cite{rowell} and Tomash\cite{tomasch}. The characteristic energy scale is given by $E_{S}^{char} = \sqrt{\frac{4 \pi^2 \mu_{S}}{x^2} +
\Delta^2}$.

Note that despite their "analogous" background, there is also a clear difference between the oscillations in $\rho_{S}$ and the Friedel oscillations. The latter are due to interfering opposite wave vectors of equal magnitude, whereas the former are caused by interference of two slightly different parallel wave vectors, $q^{e}$ and $q^{h}$.

For $E < \Delta$ (not shown in FIG. ~\ref{fig:quasi1D} ) $\rho_{S}$ is a decaying function of $x$, as single quasiparticles cannot propagate into the superconductor
(evanescent waves). The decay rate is given by $e^{-\sqrt{\Delta^2 - E^2} x/ \sqrt{\mu_{S}}}$ and the penetration depth is on the order of the superconducting coherence length\cite{hurd}, $\xi_{0} = \mu_{S}/ (k_{F} \Delta) \sim O(1000)$.
\vspace{0.5cm}

\begin{figure}
\vspace*{5mm}
\centerline{\epsfig{figure=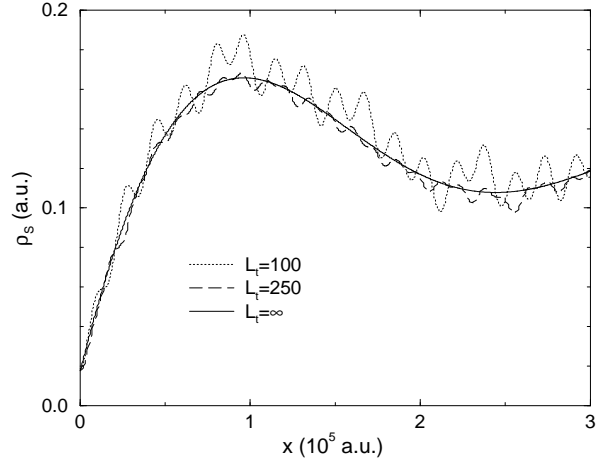,width=9cm}}
\vspace*{5mm}
\caption{ \protect{\label{fig:NS3D}}
LDOS as a function of distance $x$ in a NS junction for various transverse dimensions $L_{t}$ of the junction. The interface is located at x=0. $\mu$=0.5 Ry, $\Delta$ = 0.0001 Ry, $E/\Delta$=1.01.  
}
\end{figure}

Now we increase the transverse dimension $L_{t}$. Results in the superconductor for $E/\Delta = 1.01$ are shown in FIG.~\ref{fig:NS3D}. The summation $\sum_{n_{y}, n_{z}}$ in the LDOS with $k_{F_{x}} = \sqrt{\mu - k_{y}^2 - k_{z}^2} = \sqrt{\mu - \frac{\pi^2}{L_{t}^2}(n_{y}^2 + n_{z}^2)}$ gives rise
to fast fluctuations superimposed on the above discussed slow oscillations due to quasiparticle interference. As $L_{t}$ increases, more 
transverse modes fit into the transverse dimensions, leading to decreasing size-effects. In the limit $L_{t} \rightarrow \infty$, the summation may be replaced by an integration ($\frac{1}{L_{t}^2}
\sum_{k_{y},k_{z}} \rightarrow \frac{1}{(2\pi)^2} \int dk_{y} \int dk_{z}$)
which for the LDOS results in (for $E > \Delta$, and in AA)
\begin{eqnarray}
\rho_{S}^{\infty}\, (x,E) & \equiv & \lim_{L_{t} \rightarrow \infty} \rho_{S}\, (x,E)
\nonumber \\
& = & \lim_{L_{t} \rightarrow \infty} \frac{1}{2\pi L_{t}^2} \sum_{n_{y}, n_{z}} \left( \frac{E - \left( E - \sqrt{E^2 - \Delta^2} \right) \, {\rm cos} \left(\frac{x \sqrt{E^2 - \Delta^2}}{\sqrt{\mu_{S} - \frac{\pi^2}{L_{t}^2}(n_{y}^2 + n_{z}^2)}} \right)}{\sqrt{E^2 - \Delta^2}\, \sqrt{\mu_{S} - \frac{\pi^2}{L_{t}^2}(n_{y}^2 + n_{z}^2)}}  
\right) \nonumber \\
& = & \frac{1}{(2\pi)^2 \sqrt{E^2 - \Delta^2}} \left( E\sqrt{\mu_{S}} - 
(E - \sqrt{E^2 - \Delta^2})(\sqrt{\mu_{S}} {\rm cos} \alpha -
\sqrt{\mu_{S}} \alpha \int_{\alpha}^{\infty} \frac{{\rm sin} y}{y} dy) \right)
\label{eq:rhoSinf}
\end{eqnarray}
where $\alpha \equiv x \frac{\sqrt{E^2 - \Delta^2}}{\sqrt{\mu_{S}}}$. \\
This is the solid line in FIG.~\ref{fig:NS3D}. It is easy to verify that 
$\lim_{x \rightarrow \infty} \rho_{S}^{\infty}(E) = \frac{E \sqrt{\mu_{S}}}{(2 \pi)^2 \sqrt{E^2 - \Delta^2}}$, representing the LDOS in a 3D bulk superconductor.

In this regime of 3D NS junctions it is interesting to ask what the role of 
the AA is. As discussed in section~\ref{sec:theory}, one would expect this approximation to become worse as the number of transverse modes increases.
For the quasi-1D system the difference between values of $\rho_{S}$ with or without AA is typically $\sim 10^{-3} \%$. For systems with small $L_{t}$ ($L_{t} \sim 100$) it becomes $\sim 0.1 \%$ and for large systems at most
$\sim 1 \%$. This is a factor $10^{3}$ larger than in the 1D case,
although still not visible on the scale of FIG.~\ref{fig:NS3D}.
We are led to conclude that electrons with large transverse wavenumbers $k_{y}$ and $k_{z}$, thus with an angle of incidence deviating considerably from perpendicular to the interface, do not contribute much to the LDOS.
By using the AA, the value of the LDOS in N reduces and the size-effect fluctuations disappear, whereas in the superconductor they are both enhanced.
This can be understood by noticing that in AA the normal reflections
due to mismatch of wavevectors are neglected.
On the normal metal side of a NS junction with $\mu_{N} = \mu_{S}$, the Friedel oscillations, which are caused by normal reflections,
are thus suppressed in AA, so that $\rho_{N}$ equals $\rho_{N,\, {\rm bulk}}$. On the other hand, the oscillations in the LDOS in the superconductor, which are induced by  Andreev reflection, are enhanced in AA, due to increased quasiparticle transmission. However, since the amount of normal reflection is very small if there is no potential barrier at the interface, the enhancement of $\rho_{S}$ is also very small.

If there is a potential barrier at the interface,
then the dominant normal reflection mechanism is of course not the mismatch of wavevectors in N and S, but the presence of the barrier. In that case we would expect that application of the AA, even for large $L_{t}$, does not lead to significant changes in the value of the LDOS, at most $\sim 10^{-3} \%$.

\section{The bound state spectrum of a SNS junction}
\label{sec:SNS}

\begin{figure}[t]
%
$$\begin{picture}(150,100)
\thicklines
\put(-88,-5){\vector(0,1){50}}
\put(-92,53){E}
\put(-70,50){\line(1,0){70}}
\put(0,0){\line(1,0){100}}
\put(100,50){\line(1,0){70}}

\put(0,0){\line(0,1){50}}
\put(100,0){\line(0,1){50}}

\put(120,-15){\vector(1,0){30}}
\put(160,-17){\bf x}

\put(-3,-12){\bf 0}
\put(96,-12){\bf L}
\put(-50,30){\bf S}
\put(50,30){\bf N}
\put(150,30){\bf S}
\put(-50,65){\mbox{\boldmath $\Delta\, e^{i\phi_{L}}$}}
\put(130,65){\mbox{\boldmath $\Delta\, e^{i\phi_{R}}$}}

\end{picture}$$
\vspace*{5mm}
\caption{ \protect{\label{fig:SNSjunction}}
The planar SNS junction: the energy E and pair potential $\Delta$
are measured from the Fermi energy.}
\vspace{2cm}
\end{figure}
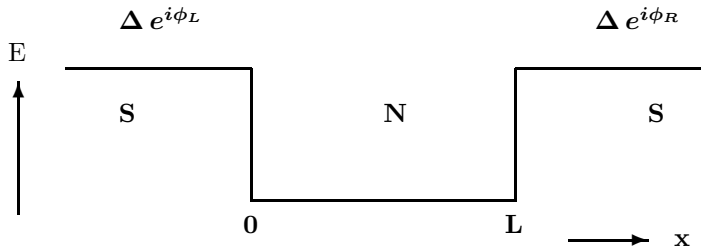

We study a SNS junction as shown in FIG.~\ref{fig:SNSjunction} with the length
$L$ of the normal region on the order of the superconducting coherence length $\xi_{0}$.
The junction has again cross section $L_{t}^2$.
The magnitude of the pair potential in both superconductors is taken equal, 
but there is a difference in phase $\delta \phi \equiv \phi_{R}-\phi_{L}$. 
For a review of this type of weak links, we refer to Likharev\cite{likharev}.

The above Josephson junction is considered with $\mu_{N} = \mu_{S} \equiv \mu$ and first in AA. So both interfaces are perfect, and there is full Andreev reflection of all quasiparticles with $E < \Delta$.

\begin{figure}
\vspace*{5mm}
\centerline{\epsfig{figure=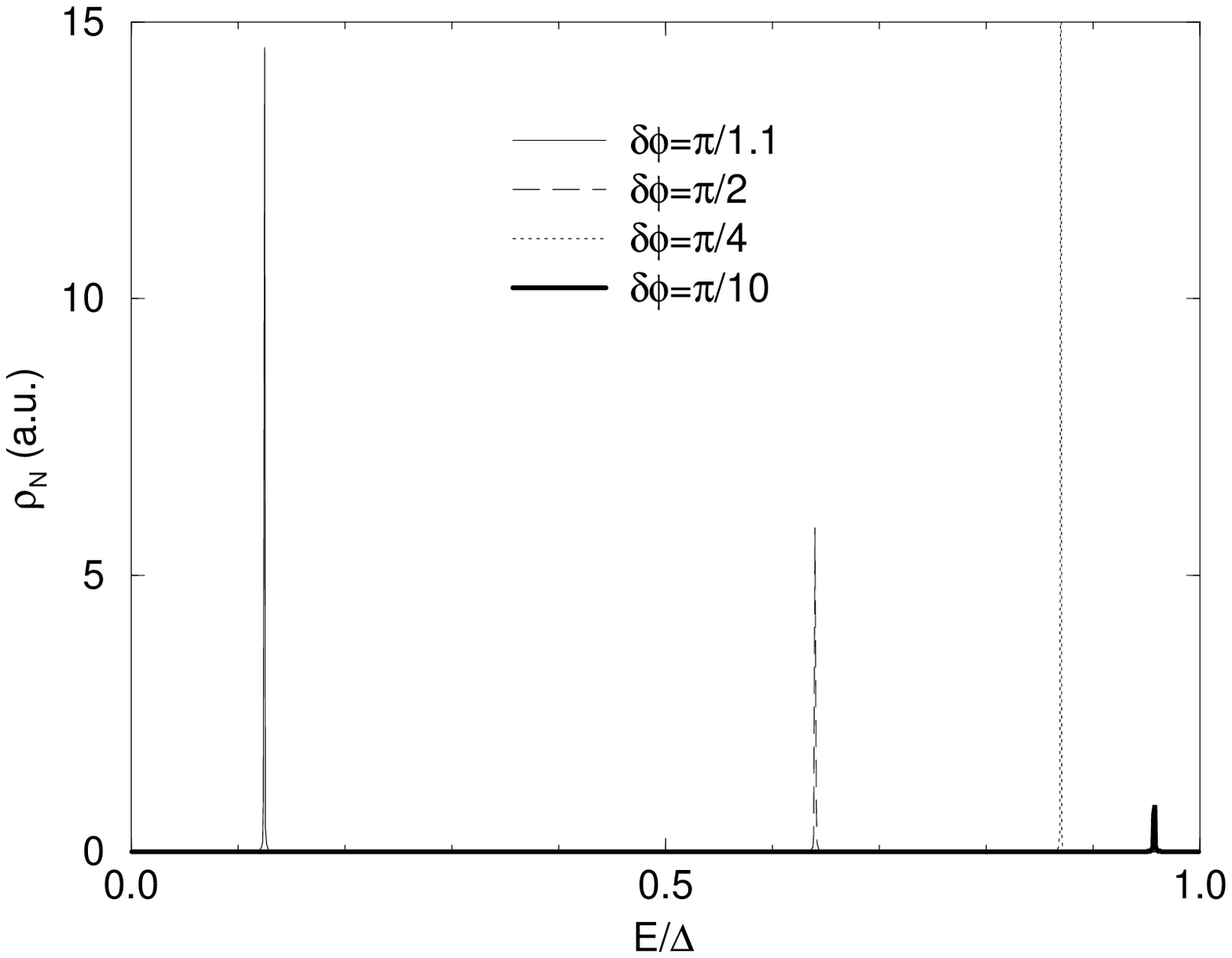,width=9cm}}
\vspace*{5mm}
\caption{ \protect{\label{fig:bound1D}}
Bound state spectrum in the normal part of a quasi-1D SNS junction, calculated {\em within} AA. $L_{t}$=4, $\mu$=0.5 Ry, $\Delta$=0.0001 Ry, $L$=2000.
}
\end{figure}

In FIG.~\ref{fig:bound1D} the LDOS in the normal region is plotted vs. $E/\Delta$, for various phase differences $\delta \phi$ and in the quasi-1D limit. The continuous spectra  for $E>\Delta$, which are proportional to
$(\mu + E)^{-1/2}$ in the one dimensional case, are not shown.
Bound states appear at energies satisfying the relation of Kulik \cite{kulik},
which was later also derived by other authors\cite{vanhouten,hurd} for a 
1D SNS junction in AA:
\be
2\pi n = 2\, \mbox{\rm arccos}\, (E/\Delta) - \frac{E}{\sqrt{\mu}}\, L \pm \delta \phi 
\label{eq:kulikrel}
\ee
In our formalism\cite{proefschrift}, Eq. (\ref{eq:kulikrel}) defines the poles 
of the scattering matrix elements in the matrix Green function; it can be understood in the following simple way. \\
An electron travelling from $x=0$ to $x=L$ acquires the phase
\[
k^{e}L - \phi_{R} -\, \mbox{\rm arccos} (E/\Delta).
\]
The first term is the phase accumulated during propagation through the normal metal; the second one is the phase shift acquired upon Andreev reflection into a hole and is equal to the phase of the pair potential in the superconductor on the right, and the third term stems from evanescently entering of the wavefunction into this superconductor\cite{beenakker3}. \\
Similarly, the back travelling hole acquires the phase
\[
-k^{h}L + \phi_{L} -\,  \mbox{\rm arccos} (E/\Delta).
\]
For constructive interference, the total phase acquired on one roundtrip should be an integer multiple of 2$\pi$, so
\[
(k^{e} - k^{h})L - 2\, \mbox{\rm arccos} (E/\Delta) +  \phi_{L} - \phi_{R}
= 2\pi n
\]
In AA, so up to terms of first order in $\frac{E}{\sqrt{\mu}}$ in the expansion of $k^{e}$ and $k^{h}$, this is the same relation as Eq.
(\ref{eq:kulikrel}).

\begin{figure}
\vspace*{5mm}
\centerline{\epsfig{figure=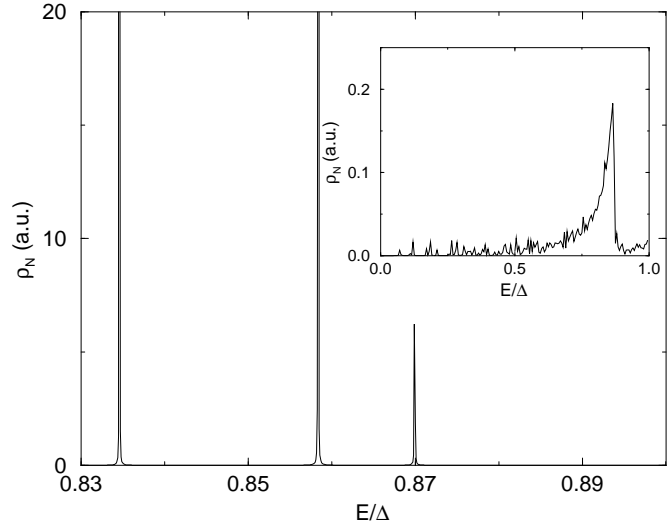,width=9cm}}
\vspace*{5mm}
\caption{ \protect{\label{fig:bound3D}}
Bound state spectrum in the normal part of a 3D SNS junction, calculated {\em within} AA. $L_{t}$=8, $\mu$=0.5 Ry, $\Delta$=0.0001 Ry, $L$=2000 and $\delta \phi$=$\pi /4$. The inset shows the same, but then for a large system with $L_{t}$=400.
}
\end{figure}

Upon increasing $L_{t}$ one obtains a picture like FIG.~\ref{fig:bound3D}, which is for 
$L_{t}$ = 8, corresponding to two transverse modes. In addition to the 1D
bound state, additional peaks appear due to the finite transverse dimensions. The width of the peaks in the figure is determined by the small imaginary part $\delta$ in the energy E+i$\delta$. The actual bound state energy is found for $\delta \to 0$. The number of peaks is equal to the number of different combinations of transverse modes in the y- and z-directions. For $L_{t}=8$,
the three peaks, from right to left, correspond to the modes with ($n_{y}$,$n_{z}$) equal to (0,0) (the 1D bound state), (1,0) or (0,1) (the bound state which corresponds to one mode in either the $y$- or $z$-direction) and (1,1) (the bound state corresponding to one mode in both the $y$- and $z$-direction).
\\
The inset of FIG.~\ref{fig:bound3D} shows that for large $L_{t}$ the discrete peaks due to the finite transverse size of the junction disappear. In the limit $L_{t} \rightarrow \infty$, one indeed expects a band\cite{ishiidos}.

All the above was done in AA. Releasing this approximation yields a bound state spectrum with a slight shift of the peaks, as compared to the same calculation in AA. Even for large junctions, this shift is $< 10 ^{-2}\,\%$ towards lower energies and we thus conclude that the AA is good.

\section{Conclusions}
\label{sec:concl}

In conlusion, we have calculated the LDOS in clean mesoscopic superconducting NS and SNS structures with finite transverse dimensions. Going from quasi-1D to 3D systems by increasing the transverse dimensions has pronounced effects on the LDOS in both types of junctions; in NS junctions additional oscillations are superimposed on the usual slow Friedel-like oscillations due to quasiparticle interference. In SNS junctions, we found additional peaks appearing in the bound state spectrum as a function of the transverse system size. 

Besides, we have tested the influence of applying the Andreev approximation, by performing all calculations of the LDOS both with and without AA. It turns out, that both in the case of a single NS interface and in the case of a SNS junction the AA does not have a large effect on the LDOS, although the AA-induced error grows by a factor of $10^{3}$ upon going from quasi-1D to 3D systems. It produces a small correction to the value of the LDOS.

Finally it is worth noting that the Green function method used is not limited to either studies of the LDOS or to the mesoscopic junctions considered here; it can also be used to study eg. supercurrents and quasiparticle currents, and it can be applied to much larger systems, such as superconducting superlattices.
The latter systems were up to now studied only\cite{tanaka} in the Andreev approximation.

\acknowledgments

We would like to thank Mr. P.W. Brouwer, Dr. K. M. Frahm and Mr. J. A. Melsen for critical comments, leading to a revised version of the manuscript. Stimulating discussions with Prof. C. van Haesendonck and Dr. A.A. Golubov are also gratefully acknowledged. Part of this work was supported by the Stichting
voor Fundamenteel Onderzoek der Materie (FOM), which is financially supported by the Nederlandse Organisatie voor Wetenschappelijk Onderzoek (NWO).


\begin{references}

\bibitem{andreev} A.F. Andreev, Zh. Eksp.Teor.Fiz. {\bf 46}, 1823 (1964) [Sov. Phys. JETP {\bf 19}, 1228 (1964)]; {\bf 51}, 1510 (1966) [{\bf 24}, 1019 (1967)]
\bibitem{mcmillan} W.L. McMillan, Phys. Rev. {\bf 175}, 559 (1968).
\bibitem{ishiidos} C. Ishii, Prog. Theor. Phys. {\bf 47}, 1464 (1972).
\bibitem{beenakker1} For a review, see C.W.J. Beenakker and H. van Houten,
Solid State Physics {\bf 44}, 1 (1991).
\bibitem{landauer} R. Landauer, IBM J. Res. Dev. {\bf 1}, 223 (1957).
\bibitem{buttiker} M. B\"{u}ttiker, Y. Imry, R. Landauer and S. Pinhas,
Phys. Rev. B {\bf 31}, 6207 (1985).
\bibitem{lambert} C.J. Lambert, J. Phys. Cond. Matter {\bf 3}, 7331 (1991).
\bibitem{beenakker2} C.W.J. Beenakker, Phys. Rev. Lett. {\bf 67}, 3836 (1991).
\bibitem{akkermans} E. Akkermans, A. Auerbach, J.E. Avron and B. Shapiro,
Phys. Rev. Lett. {\bf 66}, 76 (1991).
\bibitem{furusaki} A. Furusaki and M. Tsukada, Phys. Rev. B {\bf 43}, 10164 (1991).
\bibitem{tanaka} Y. Tanaka and M. Tsukada, Phys. Rev. B {\bf 44}, 7578 (1991).
\bibitem{kiesplehn} G. Kieselmann, Phys. Rev. B {\bf 35}, 6762 (1987); H. Plehn, O.-J. Wacker and R. K\"{u}mmel, Phys. Rev. B {\bf 49}, 12140 (1994).
\bibitem{proefschrift} R.T.W. Koperdraad, {\em  The Proximity Effect in Superconducting Metallic Multilayers}, PhD thesis Vrije Universiteit, Amsterdam, 1995, available on request.
\bibitem{rutger} R.T.W. Koperdraad, A. Lodder, in preparation.
\bibitem{ishii} C. Ishii, Prog. Theor. Phys. {\bf 44}, 1525 (1970).
\bibitem{rowell} J.M. Rowell and W.L. McMillan, Phys. Rev. Lett. {\bf 16}, 453 (1966); W.L. McMillan and J.M. Rowell in {\em Superconductivity} Vol. I, edited by R.D. Parks (Dekker, New York, 1969).
\bibitem{tomasch} W.J. Tomasch, Phys. Rev. Lett. {\bf 16}, 16 (1966); 
{\bf 15}, 672 (1965).
\bibitem{crommie} M.F. Crommie, C. P. Lutz and D. M. Eigler, Nature {\bf 363}, 524 (1993).
\bibitem{degennes} P.G. de Gennes, {\em Superconductivity of Metals and Alloys},
Benjamin, New York, 1966.
\bibitem{hurd} M. Hurd and G. Wendin, Phys. Rev. B {\bf 49}, 15258 (1994).
\bibitem{likharev} K.K. Likharev, Rev. Mod. Phys. {\bf 51}, 101 (1979).
\bibitem{kulik} I.O. Kulik, Zh. Eksp. Teor. Fiz. {\bf 57}, 1745 (1969) [Sov. Phys. JETP {\bf 30}, 944 (1970)].
\bibitem{vanhouten} H. van Houten and C.W.J. Beenakker, Physica B {\bf 175}, 
187 (1991).
\bibitem{beenakker3} C.W.J. Beenakker, Phys. Rev. B {\bf 46}, 12841 (1992).


\end{references}
\end{document}